\documentstyle[12pt,aps]{revtex}
\begin{document}
\draft
\preprint{June 1996; DART-HEP-96/03}

\title{Matching numerical simulations to continuum field
theories: A lattice renormalization study}

\author{Julian Borrill and Marcelo Gleiser\footnote{NSF Presidential 
Faculty Fellow}\footnote{email: borrill@nevill.dartmouth.edu, 
gleiser@peterpan.dartmouth.edu }}
\address{ Department of Physics and Astronomy, Dartmouth College,
Hanover, NH 03755 }

\date{\today}
\maketitle
\begin{abstract}
The study of nonlinear phenomena in systems with many degrees of
freedom often relies on complex numerical simulations. In trying
to model realistic situations, these systems may be coupled to
an external environment which drives their dynamics. For
nonlinear field theories coupled to thermal (or quantum) baths,
discrete lattice formulations must be dealt with extreme care if
the results of the simulations are to be interpreted in the
continuum limit. Using techniques from renormalization theory, a
self-consistent method is presented to match lattice results to
continuum models. As an application, symmetry restoration in
$\phi^4$ models is investigated.

\end{abstract}


\newpage

\baselineskip 24pt

\noindent
The study of nonlinear phenomena has changed dramatically during
the last two decades or so, as an increasing number of once
forbidding problems have become amenable to treatment by faster
and cheaper computers. From coupled anharmonic oscillators to
gravitational clustering, from plasma physics to the dynamics of
phase transitions, numerical simulations are often the only tool
to probe the physics of complex nonlinear systems \cite{Nonlin}.

Typically, we are interested in investigating the behavior of a
particular physical system described by ordinary or partial
nonlinear differential equations. In the present work, focus
will be mostly on the latter case, which can be thought of as
representing systems with finitely or infinitely many coupled
degrees of freedom.  Apart from very few exceptions, such as
kink solutions for sine-Gordon or $\phi^4$ models
\cite{Solitons}, nonlinear partial differential equations have
no analytical solutions. The situation is even worse if we
attempt to model realistic behavior by coupling the system to an
external environment. This external environment often represents
a thermal or quantum bath, adding an element of stochasticity to
the deterministic evolution equations. In order to gain some
insight into the role of nonlinearities, perturbation theory is
frequently used. However, examples ranging from the simple
pendulum equation \cite{Pendulum} to critical phenomena during
phase transitions \cite{Crit-phen} remind us that perturbation
theory breaks down precisely in the region of parameter space
where nonlinear effects become predominant.

The alternative is to address the problem numerically, solving
the equations of interest using a computer. In the case of
partial differential equations, the problem is set up on a
lattice which represents a particular choice of discretization
procedure.  For a function of $d$-dimensional
position and time, $f({\bf x},t)$, satisfying some partial
differential equation with given initial and boundary
conditions, we typically construct a $d$-dimensional lattice of a
given geometry, say cubic or triangular, to represent space at a
particular instant, and replicate it at (usually regular)
intervals to represent time. The continuous function may then be
discretized following well-prescribed rules by which continuous
derivatives are approximated by finite ratios of the lattice
variables \cite{Num-Rec}.

The use of a spatial lattice introduces two artificial length
scales; the `macroscopic' size of the lattice in each dimension,
$L$, and the `microscopic' distance between neighbouring lattice
points, $\delta x$. These length scales provide bounds on the
wavelengths of modes which can be represented on the lattice,
whilst the total the number of lattice points $N$ (for cubic
lattices being $N = (L/\delta x)^d$) is the restricted number of
degrees of freedom being integrated at each time step.
Computational physicists (and computers) spend a considerable
amount of time trying to get around the limitations that these
length scales introduce to numerical studies of continuum
systems. Occasionally, one or other of these limitations may
become insignificant due to the particular physical behavior of
the system; for example, close to the critical point of a second
order phase transition the divergence of the characteristic
length scale of the system means that its bulk properties (and
in particular its critical exponents) are determined by the long
wavelength modes alone, doing away with the need for the high
spatial resolution given by a small lattice spacing $\delta x$
\cite{Crit-phen}. In general, however, since the continuum
corresponds to the limit $L \rightarrow \infty, \delta x
\rightarrow 0, N \rightarrow \infty$, a better approximation is
obtained from a larger and finer lattice, leading to the notion
of the continuum limit of a discrete system. For continuum
systems described by continuous functions, such as fluids,
fields, or deformable bodies, our discrete representation should
have a well-defined continuum limit, {\it i.e.}, one that is
stable as $\delta x\rightarrow 0$ (at fixed $L$). Moreover, we
should also demand that it is a {\it good} continuum limit, in
that it matches the original continuum system. As discussed
below, for systems coupled to external environments, even if the
continuum limit can be achieved on the lattice it is not always
clear how to match the lattice results to a continuum
theory. These two questions --- how to achieve a continuum limit
in lattice simulations, and how to ensure that it is a good
limit, in the sense of matching the appropriate continuum theory
--- are the focus of this work.

For linear systems, achieving a continuum limit does not usually
present any difficulties. Typically there is a minimal
length-scale in the problem which can be used as a guideline for
the choice of $\delta x$. For example, when solving the wave
equation, it is possible to find a small enough $\delta x$ and
show that the same results are obtained if smaller values are
used, provided one makes sure the discretization of time is
appropriately chosen so that the evolution is stable.

For nonlinear systems, the situation is more complicated. If we
think for a moment in terms of a Fourier decomposition of the
function $f({\bf x},t)$, the effect of nonlinearities is to
couple different wavelength modes in a nontrivial way; the
dynamics of short wavelength modes will influence the dynamics
of long wavelength modes and {\it vice-versa}. Mechanisms to
handle this problem are sensitive both to the particular system
under study and to which of its properties are of interest,
often seeming to be more an art than a science. For example, if
we are solely interested in the dynamics of long wavelength
modes with slow relaxation time-scales, it may be possible to
add extra artificial terms to the evolution equations which damp
the behavior of faster modes. For situations in which nonlinear
fields are coupled to an external environment with stochastic
properties, say a thermal (or quantum) bath, a detailed
investigation of how to approach the continuum limit on the
lattice is lacking. This does not imply that this problem has
been completely overlooked, but that it may have received less
attention than it deserves.

In the context of classical field theories at finite temperature
there has been some work on obtaining such a continuum limit.
For example, Parisi \cite{Parisi} suggested the addition of
renormalization counterterms, a proposal then implemented by
Alford and Gleiser \cite{AG} in the context of 2-dimensional
nucleation studies (albeit with a somewhat {\it ad hoc} match to
a continuum theory), and by Kajantie {\it et al.}
\cite{Shaposhnikov} in lattice gauge simulations of the
electroweak phase transition. Alford and Gleiser in particular
showed that neglecting lattice spacing effects in the numerical
determination of nucleation rates can lead to severe errors in
the measured values. This conclusion is not particular to
systems exhibiting metastable states, but to any nonlinear field
model in contact with external stochastic environments. Thus,
the issues that are raised here are of concern to a wide range
of physical systems modelled through the separation of system
and environment, from quantum field theories to effective field
theories describing condensed matter systems.

Even if a continuum limit can be achieved on the lattice, we
must still ensure that the numerical results correspond to the
appropriate continuum theory. In general, the coupling to a
stochastic environment modifies the effective lattice theory,
which cannot be naively matched to the original continuum
model. The question then becomes what theory is the lattice
simulating, and can we extract it in a self-consistent way?
These questions will be addressed below in the context of two
continuous nonlinear models in 2+1 dimensions, one temperature
independent and the other temperature dependent (the well-known
Ginzburg-Landau model). Both models describe phase transitions
in the Ising universality class. Extensions to $d+1$ dimensions
should be straightforward.

\vspace{0.5cm}

\noindent
{\bf Formulating continuum models on a lattice: The issues}

Consider a single scalar field $\phi({\bf x}, t)$ in a potential
$V_0(\phi)$ which may or not be temperature dependent. This
potential can model interactions of $\phi$ with itself and with
other fields.  For example, a linear term of the form $\phi
{\cal H}$ is often used to represent the coupling of $\phi$ to
an external magnetic field for models of ferromagnetic
transitions. In this report, focus will be on potentials which
are simple polynomials of even power in $\phi$, although our
approach is equally valid for potentials with odd powers of
$\phi$, typical of nucleation studies.  The Hamiltonian for this
system is, (in units of $c=k_B=1$)
\begin{equation}\label{e:hamiltonian}
{{H[\phi]}\over {T}}={1\over {T}}\int d^2x\left [{1\over 2}\left (
\nabla \phi \cdot \nabla \phi \right )+V_0(\phi)\right ]~.
\end{equation}

The field $\phi$ can be thought of as representing a scalar
order parameter in models of phase transitions in the Ising
universality class, such as ferromagnets, binary fluid mixtures,
metal alloys, or in studies of domain wall formation in
cosmology. As such, it is convenient to model its dynamics in
contact with a heat bath by means of a generalized Langevin
equation,
\begin{equation}\label{e:langevin}
{\partial^2\phi\over\partial t^2} = \nabla^2\phi - \eta 
{\partial\phi\over\partial t}
- {\partial V_0 \over \partial \phi} + \xi({\bf x},t)~~,
\end{equation}
where the viscosity coefficient $\eta$ is related to the
stochastic force of zero mean $\xi({\bf x},t)$ by the
fluctuation-dissipation relation,
\begin{equation}\label{e:fluct-diss}
\langle \xi({\bf x}, t) \xi({\bf x}', t') \rangle = 2 \eta T
\delta({\bf x} - {\bf x}') \delta(t - t')~.
\end{equation}
This approach guarantees that $\phi$ will be driven into
equilibrium, although the time-scale $\eta^{-1}$ is
arbitrary. It has been extensively used in numerical simulations
of thermal creation of kink-antikink pairs \cite{Kinks},
nucleation \cite{AG,Num-nuc}, spinodal decomposition
\cite{Spin-dec}, and pattern formation in the presence of
external noise \cite{Pattern}, to mention but a few examples.
Note that in the high viscosity limit the second-order time
derivative can be neglected, as is common practice in systems
with slower dynamical time-scales.

The next step is to discretize this system and cast it on a
lattice. Using a standard second-order staggered leapfrog method
we can write,
\begin{eqnarray}\label{e:lattice_equation}
\dot{\phi}_{i,m+1/2} & = & \frac{(1 - \frac{1}{2} \eta \delta t)
\dot{\phi}_{i,m-1/2} + \delta t (\nabla^{2} \phi_{i,m} -
V'_0(\phi_{i,m}) + \xi_{i,m})}{1 + \frac{1}{2} \eta \delta t}
\nonumber \\
\phi_{i,m+1} & = & \phi_{i,m} + \delta t \dot{\phi}_{i,m+1/2}
\end{eqnarray}
where $i$-indices are spatial and $m$-indices temporal, overdots
represent derivatives with respect to $t$ and primes with
respect to $\phi$. The discretised fluctuation-dissipation
relation now reads
\begin{equation}
\langle \xi_{i,m} \xi_{j,n} \rangle = 2 \eta T
\frac{\delta_{i,j}}{\delta x^{2}} \frac{\delta_{m,n}}{\delta t}
\end{equation}
so that
\begin{equation}
\xi_{i,m} = \sqrt{\frac{2 \eta T}{\delta x^{2} \delta t}} G_{i,m}
\end{equation}
where $G_{i,m}$ is taken from a zero-mean unit-variance Gaussian.

Note that as a first guess we have used $V_0(\phi)$ in the
lattice formulation of the model. Is this the correct procedure?
It is well-known that classical field theory in more than one
spatial dimension is ultraviolet divergent, the Rayleigh-Jeans
ultraviolet catastrophe \cite{UVcatas}. Formulating the theory
on a lattice takes care of the problem, as a sharp momentum
cutoff is introduced by the lattice spacing $\delta x$, with
$\Lambda=\pi/\delta x$.  However, a finite lattice spacing
creates two difficulties. First, the lattice theory is
coarse-grained on the scale $\delta x$; in other words, the
lattice theory is not equivalent to the continuum theory we
started with, and our results will depend on $\delta x$, unless
this dependence is handled by a proper renormalization
procedure. Second, if the lattice theory is not equivalent to
the continuum theory we started with, to what continuum theory
is it equivalent to?  Fortunately, there is a well-defined
procedure that addresses both difficulties at once. Within its
validity, it is possible to establish a one-to-one
correspondence between lattice simulations and field theories in
contact with stochastic baths.

\vspace{0.5cm}

\noindent
{\bf Formulating continuum models on a lattice: The procedure}

In order to recover the continuum limit on the lattice we must
eliminate any dependence on the cutoff. The coupling to the heat
bath will induce fluctuations on all possible scales.  Since the
cutoff sets the scale for the smallest possible spatial
fluctuations in the system, we may incorporate the effects of
all fluctuations down to the smallest scale using perturbation
theory. Thus, the lattice theory must be equivalent to a
continuum theory with a sharp ultraviolet cutoff.  For classical
field theories, the one-loop corrected effective potential with
a large momentum cutoff is given by \cite{CFT},
\begin{equation}
V_{\rm 1L}(\phi)=V_0 + {T\over 2}\int_0^{\Lambda}{{d^2p}\over
{(2\pi)^2}}{\rm ln}\left (p^2 + V_0''\right ) +{\rm counterterms}~.
\end{equation}
These theories describe fluctuations with $\hbar \omega \ll
k_BT$. In semi-classical language, the excitations of the
field contain many fundamental quanta.  Note that there is a
one-to-one correspondence between classical statistical field
theory in $d+1$ dimensions and Euclidean quantum field theory in
$d$ dimensions. While the loop expansion is in powers of $T$ for
the former, it is in powers of $\hbar$ for the latter. For
$d=2$, the only divergences are at one loop, although higher
loops can generate finite terms which modify the effective
Hamiltonian. The dependence on the cutoff $\Lambda$ can be
handled by introducing proper counterterms.

Integration gives,
\begin{equation}
V_{\rm 1L}(\phi)= V_0 + \frac{T}{8 \pi} V''_0 \left [1 - \ln
\left(\frac{V''_0}{\Lambda^{2}}\right)\right ] + {\rm
counterterms}~.
\end{equation}
The form of $V_0$ will determine the counterterms needed to
cancel the dependence on $\Lambda$. For polynomial potentials of
order $\phi^n$, one typically needs counterterms up to order
$\phi^{n-2}$. In the case of interest here, degenerate
double-well potentials, only one quadratic counterterm is
needed, of form $a\phi^2$, with $a$ constant.  As usual, the
value of $a$ is fixed by imposing a renormalization condition.
Because of the logarithmic divergence, the renormalization
condition must be imposed at some energy scale $M$, which is
chosen to be,
\begin{equation}
V''_{\rm 1L}(\phi = \sqrt M) = V''_{\rm 0}(\phi = \sqrt M)~.
\end{equation}
The renormalized one-loop corrected potential is then,
\begin{equation}\label{e:oneloop}
V_{\rm 1L}(\phi) = V_{\rm 0} + \frac{T}{8 \pi} V''_{\rm 0} \left
[1 - \ln \left (\frac{V''_{\rm 0}}{\Lambda^{2}}\right )\right ]
+ \frac{T}{16 \pi} \left (V''''_{\rm 0} \ln \left. \left
(\frac{V''_{\rm 0}}{\Lambda^{2}}\right ) + \frac{(V'''_{\rm
0})^{2}}{V''_{\rm 0}}\right )\right |_{\phi = \sqrt M} \phi^{2}
\end{equation}

The above procedure incorporates thermal fluctuations to the
original potential $V_0(\phi)$ at some energy scale $M$ to
one-loop order.  As with any perturbative approach, it will
break down wherever large amplitude fluctuations are present,
and in particular close to the critical point $T_c$. Although
there are techniques to improve the perturbative expansion in
the neighborhood of the critical point, such as
$\varepsilon$-expansion methods \cite{Varepsilon} (not too
reliable for 2-d), in this work we will concentrate on the
matching of the continuum theory to the lattice simulation in
regions of the parameter space where the one-loop calculation is
valid. Close to criticality the theory of Eq. \ref{e:oneloop}
breaks down, and we restrict our investigation to the extraction
of the critical exponent controlling the divergence of the order
parameter.

How is this continuum theory matched to the lattice simulation?
The procedure we propose is quite simple. Since the continuum
theory above incorporates fluctuations from momentum scales up
to $\Lambda$, we write the lattice potential as,
\begin{equation}
V_{\rm latt}(\phi) = V_0 + a\phi^2~,
\end{equation}
where $a$ is fixed by the renormalization condition in the
continuum, but with $\Lambda = \pi/\delta x$. That is,
\begin{equation}
\label{e:goodcontlimit}
V_{\rm latt}(\phi) = V_0 + \frac{T}{16 \pi} \left (V''''_{\rm 0}
\ln \left. \left (\frac{V''_{\rm 0}}{(\pi/\delta x)^{2}}\right )
+ \frac{(V'''_{\rm 0})^{2}}{V''_{\rm 0}}\right )\right |_{\phi =
\sqrt M}\phi^2~.
\end{equation}
As we show below, this procedure takes care of the two problems
raised by formulating the continuum theory on the lattice,
namely, the dependence of lattice results on lattice spacing and
the matching of the lattice theory to the continuum at some
renormalization energy scale $M$. The generic emergence of a
good continuum limit from Eq. \ref{e:goodcontlimit} is
the central result of this work.

\vspace{0.5cm}

\noindent
{\bf Applications}

We will apply the above procedure to two cases, with potentials
which are temperature independent and temperature dependent,
respectively.  Consider first the temperature-independent
potential,
\begin{equation}
V_{\rm 0}(\phi) = - \frac{1}{2} m^{2} \phi^{2} + \frac{1}{4}
\lambda \phi^{4}~.
\end{equation}
Choosing the renormalization point to be 
$\phi_{RN} = \sqrt \frac{M^{2} + m^{2}}{3 \lambda}$, the renormalized
continuum potential is, from Eq. \ref{e:oneloop},
\begin{equation}\label{e:renorm-cont}
V_{\rm 1L}(\phi) = - \frac{1}{2} m^{2} \phi^{2} + \frac{1}{4}
\lambda \phi^{4} + \frac{3 \lambda T}{8 \pi} \left( 1 + 2 \;
\frac{M^{2} + m^{2}}{M^{2}} \right) \phi^{2} - \frac{T}{8 \pi}
(3 \lambda \phi^{2} - m^{2}) \ln \left( \frac{3 \lambda \phi^{2}
- m^{2}}{M^{2}} \right)~.
\end{equation}
It is convenient to introduce dimensionless variables (because
there is no $\hbar$ in this theory, $m$ has dimensions of
(length)$^{-1}$ while $\phi$ has dimensions of
(energy)$^{1/2}$), $\tilde{x} = x m, \; \tilde{t} = t m, \;
\tilde{\phi} = \phi \lambda^{1/2} m^{-1}, \; \tilde{\eta} = \eta
m^{-1}, \; \theta = T \lambda m^{-2}, \; \tilde{M} = M m^{-1},
\; \tilde{\Lambda} = \Lambda m^{-1}$. From the discussion in the
previous section, the lattice-spacing independent lattice
potential is, using dimensionless variables (and dropping the
tildes),
\begin{equation}
V_{\rm latt}(\phi) = - \frac{1}{2} \phi^{2} + \frac{1}{4}
\phi^{4} + \frac{3 \theta}{4 \pi} \left( \ln \left( \frac{M
\delta x}{\pi} \right) + \frac{M^{2} + 1}{M^{2}} \right)
\phi^{2}~.
\end{equation}
Fig. 1 shows the impact of the added counterterm to the lattice
results. We display the time evolution of the spatially averaged
field, ${\bar \phi}={1\over A}\int dA \phi$, starting from a
broken symmetric phase ${\bar \phi}=-1$, without the counterterm
(Fig. 1a) and with the counterterm (Fig. 1b). The parameters
$\theta,~ M$, and physical lattice size $L$, were kept fixed,
and only the lattice spacing $\delta x$ was varied. (Throughout
this work we keep the viscosity coefficient $\eta=1$ as we are
only interested in final equilibrium quantities.) Clearly,
omitting the counterterm leads to severe lattice spacing
dependence of the results, even to the point of having symmetry
restoration. Experiments varying $\theta$ and $M$ showed that
the procedure is robust, with excellent $\delta x$-independence
being achieved, even close the critical point, as long as the
expansion parameter $\theta/8\pi \ll 1$.

The next step is to compare the lattice results with the
continuum models of Eq. \ref{e:renorm-cont} in their domain of
validity. Being perturbative, we expect the continuum models to
break down when the fluctuations become large, at high
temperatures or close to the critical point. By contrast, the
lattice models incorporate fluctuations up to the limiting size
$L$, and so may remain valid even when the continuum models
break down. The continuum potential gives a prediction for the
critical temperature of
\begin{equation}
\theta_{c} = \frac{2 \pi}{3 (1 + M^{-2} + \ln M)}~.
\end{equation}
Note that $\theta_c$ has its maximum value at $M^2 = 2$; as we
move away from this point in either direction $\theta_c$
decreases, and we should expect perturbation theory to continue
to be a valid approximation closer and closer to the critical
point. Ultimately, however, the phase transition is
nonperturbative, the field fluctuations become large, and
perturbation theory must fail. Fig. 2 shows the variation in the
equilibrium mean field value $\bar{\phi}_{\rm eq}$ with temperature
$\theta$, squares from the lattice and lines from the continuum,
for values of the renormalization energy-scale $M=0.1$
(Fig. 2a), $M=\sqrt{2}$ (Fig. 2b), and $M=10$ (Fig. 2c). The
discontinuities in the continuum are related to the concavity of
the corrected potential between the inflection points, which
gives rise to an imaginary part. As shown by Weinberg and Wu
\cite{WW}, the imaginary part of the potential represents
unstable physical states typical of the process of phase
separation; the figure shows only the real part of the corrected
potential. There is indeed excellent agreement at low
temperatures, which is progressively lost as the temperature
increases.

At the one-loop level, perturbation theory is equivalent to mean
field theory.  Close to the critical point, where mean field
theory breaks down, we expect the equilibrium value of ${\bar
\phi}$ to diverge as a power law,
\begin{equation}
\bar{\phi}_{\rm eq} \propto \left( \frac{\theta_{c} -
\theta}{\theta_{c}} \right)^{\beta}
\end{equation}
with the critical exponent $\beta=1/2$ for mean field theory and
$\beta=1/8$ for the 2-d Ising model.  Figure 3 shows the
behavior of the lattice and continuum equilibrium mean field
values $\bar{\phi}_{\rm eq}$ with reduced temperature $\theta_{r} \equiv
(\theta_{c} - \theta)/\theta_{c}$ for $M = 0.1$ --- squares
being results from the lattice simulations, triangles the
predicted behavior from the continuum, and the lines indicating
the two slopes $\beta =1/2$ and $\beta =1/8$. We see that the
continuum perturbation theory behaves as a mean field theory,
whilst the lattice theory in the neighborhood of the critical
point is in the universality class of the 2-d Ising model as
expected.

We now consider the case of a temperature-dependent potential.
The goal is to show that the above procedure works equally well
in this case; both lattice-spacing independence and the matching
to a continuum theory can be achieved in a consistent way.
Coupling a temperature-dependent potential to a heat bath does
not necessarily imply a double counting of the thermal degrees
of freedom. The choice of potential $V_0$ simply reflects
different physical models. For example, one may include
phenomenological temperature-dependent terms in $V_0$, as in the
Ginzburg-Landau model, or may obtain temperature corrections by
integrating out from the partition function either other fields
coupled to $\phi$ or short wavelength modes of the field $\phi$
itself \cite{GR}. In either case, the heat bath may then be
representing stochastic forces not included in the integration
process, or simply an external environment coupled to $\phi$
phenomenologically, which drives the system to its final
equilibrium state. As an example, we choose the Ginzburg-Landau
potential,
\begin{equation}
V_{\rm 0}(\phi) = \frac{1}{2} a (T - T_{c}') \phi^{2} + \frac{1}{4}
\lambda \phi^{4}~,
\end{equation}
where the prime is a reminder that the critical temperature has
an arbitrary value in the mean field model. Fixing the
renormalization energy scale at $\phi_{RN} = \sqrt \frac{M^{2} -
a (T - T_{c}')}{3 \lambda}$, the renormalized continuum
potential becomes,
\begin{eqnarray}
V_{\rm 1L}(\phi) & = & \frac{1}{2} a (T - T_{c}') \phi^{2} +
\frac{1}{4} \lambda \phi^{4} + \frac{3 \lambda T}{8 \pi} \left (
1 + 2 \; \frac{M^{2} - a (T - T_{c}')}{M^{2}} \right ) \phi^{2}
\nonumber \\ && - \frac{T}{8 \pi} \left [3 \lambda \phi^{2} + a
(T - T_{c}')\right ] \ln \left ( \frac{3 \lambda \phi^{2} + a (T
- T_{c}')}{M^{2}} \right )~.
\end{eqnarray}
Following the same steps as before and arbitrarily setting
$\theta_{c}' = 1$, this theory is matched on the lattice to
\begin{equation}
V_{\rm latt}(\phi) = \frac{1}{2} (\theta - 1) \phi^{2} +
\frac{1}{4} \phi^{4} + \frac{3 \theta}{4 \pi} \left [ \ln \left(
\frac{M \delta x}{\pi} \right) + \frac{M^{2} - (\theta - 1)}{M^{2}} 
\right ] \phi^{2}~.
\end{equation}
Fig. 4 compares the lattice results without (Fig. 4a) and with
(Fig. 4b) the renormalization counterterm. The prescription to
obtain lattice-spacing independence works equally well in this
case. Fig. 5 again compares the lattice simulations (squares)
and the continuum model (lines) for renormalization scales
$M=0.1$ (Fig. 5a), $M=\sqrt{2}$ (Fig. 5b), and $M=10$
(Fig. 5c). For low temperatures excellent agreement is obtained,
as in the temperature independent case. Note that this also
confirms that our model has not been `twice-cooked'; had it
been, no such agreement would be possible. Finally, in Fig. 6,
we show the critical behavior of the lattice (squares) and
continuum (triangles) for $M=0.1$. Again the lattice obtains the
Ising critical exponent, $\beta=1/8$, close to criticality.

In summary, we have presented a self-consistent method to match
lattice simulations to nonlinear field theories in contact with
an external stochastic environment. This approach is of
potential interest in a wide range of physical problems, from
noise-induced pattern-forming instabilities and phase separation
in condensed matter physics to symmetry breaking in high energy
physics and cosmology. It was shown that adding the right
renormalization counterterms to the lattice potential provides a
good continuum limit, independent of the lattice-spacing and
matching the appropriate continuum theory.  That this matching
breaks down at high temperatures and/or close to a critical
point is not surprising, as it reflects the limitations of
perturbation theory in probing critical phenomena
quantitatively. The procedure was demonstrated to work well for
a large class of widely-used potentials --- both temperature
independent and dependent --- and over a wide range of the
renormalization energy scale $M$.

\acknowledgements
Julian Borrill was supported by a National Science Foundation
grant no. PHY-9453431. Marcelo Gleiser was partially supported
by the National Science Foundation
through a  Presidential Faculty Fellows
Award no. PHY-9453431 and by a National Aeronautics and Space Administration
grant no. NAGW-4270.

\listoffigures

Figure 1. The time evolution of the mean field $\bar{\phi}(t)$
at five different lattice spacings $\delta x = 0.125, \, 0.25,
\, 0.5, \, 1.0$ and $2.0$ for the temperature independent
potential --- (a) without the renormalisation counterterms added
($\delta x$ increasing downwards), and (b) with the
renormalisation counterterms added.\\

Figure 2. The variation in the equilibrium mean field
$\bar{\phi}_{\rm eq}$ with the dimensionless temperature $\theta$
from the lattice (squares) and the continuum (lines) for the
temperature independent potential --- (a) for $M=0.1$, (b) for
$M = \sqrt 2$, and (c) for $M = 10$.\\

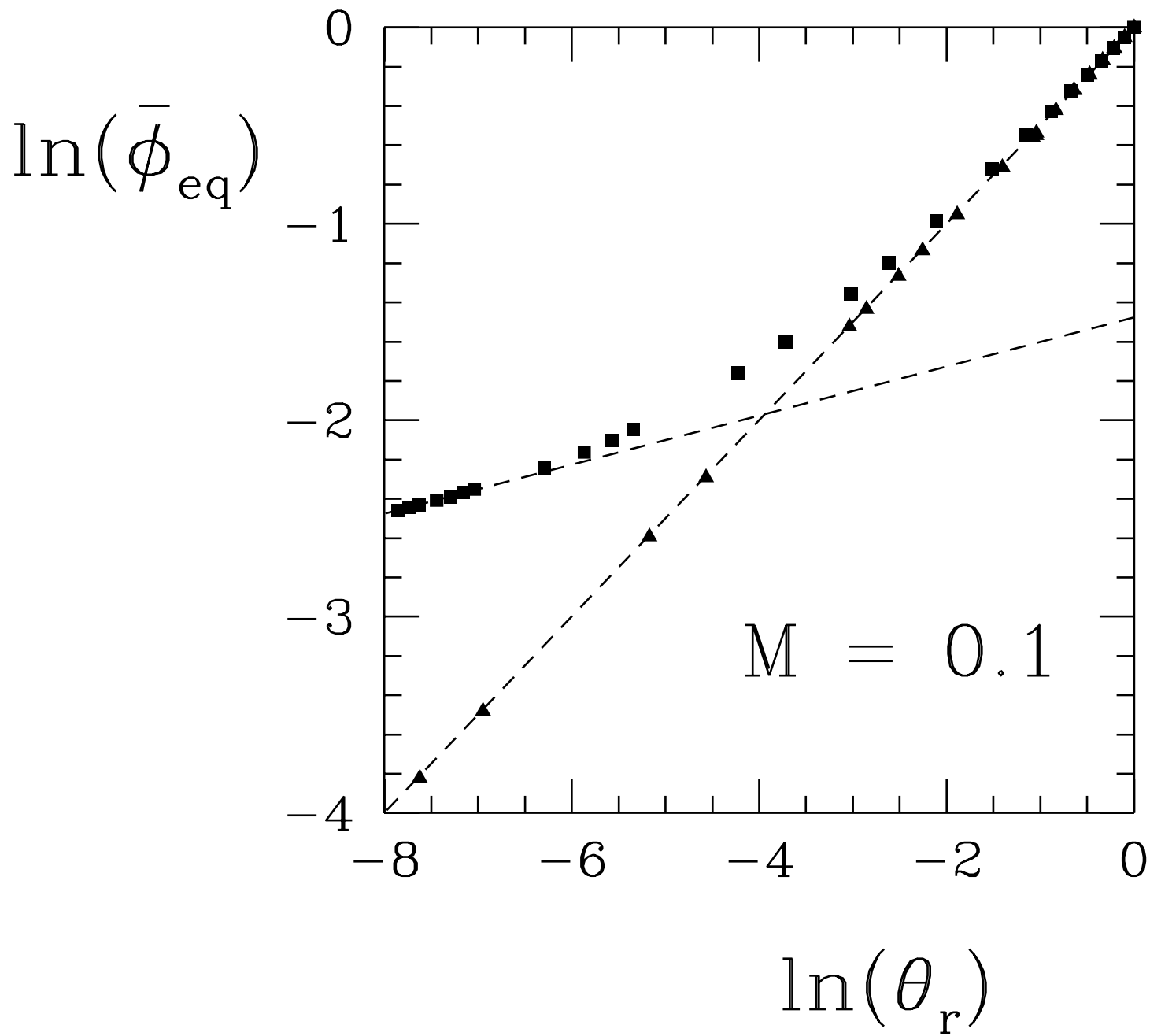
Figure 3. The variation in the equilibrium mean field
$\bar{\phi}_{\rm eq}$ with the reduced dimensionless temperature
$\theta_{r}$ from the lattice (squares) and the continuum
(triangles) for the temperature independent potential. The
dashed lines have slopes of $1/8$ and $1/2$.\\

Figure 4. The time evolution of the mean field $\bar{\phi}(t)$
at five different lattice spacings $\delta x = 0.125, \, 0.25,
\, 0.5, \, 1.0$ and $2.0$ for the temperature dependent
potential --- (a) without the renormalisation counterterms added
($\delta x$ increasing downwards), and (b) with the
renormalisation counterterms added.\\

Figure 5. The variation in the equilibrium mean field
$\bar{\phi}_{\rm eq}$ with the dimensionless temperature $\theta$
from the lattice (squares) and the continuum (lines) for the
temperature dependent potential --- (a) for $M=0.1$, (b) for $M
= \sqrt 2$, and (c) for $M = 10$.\\

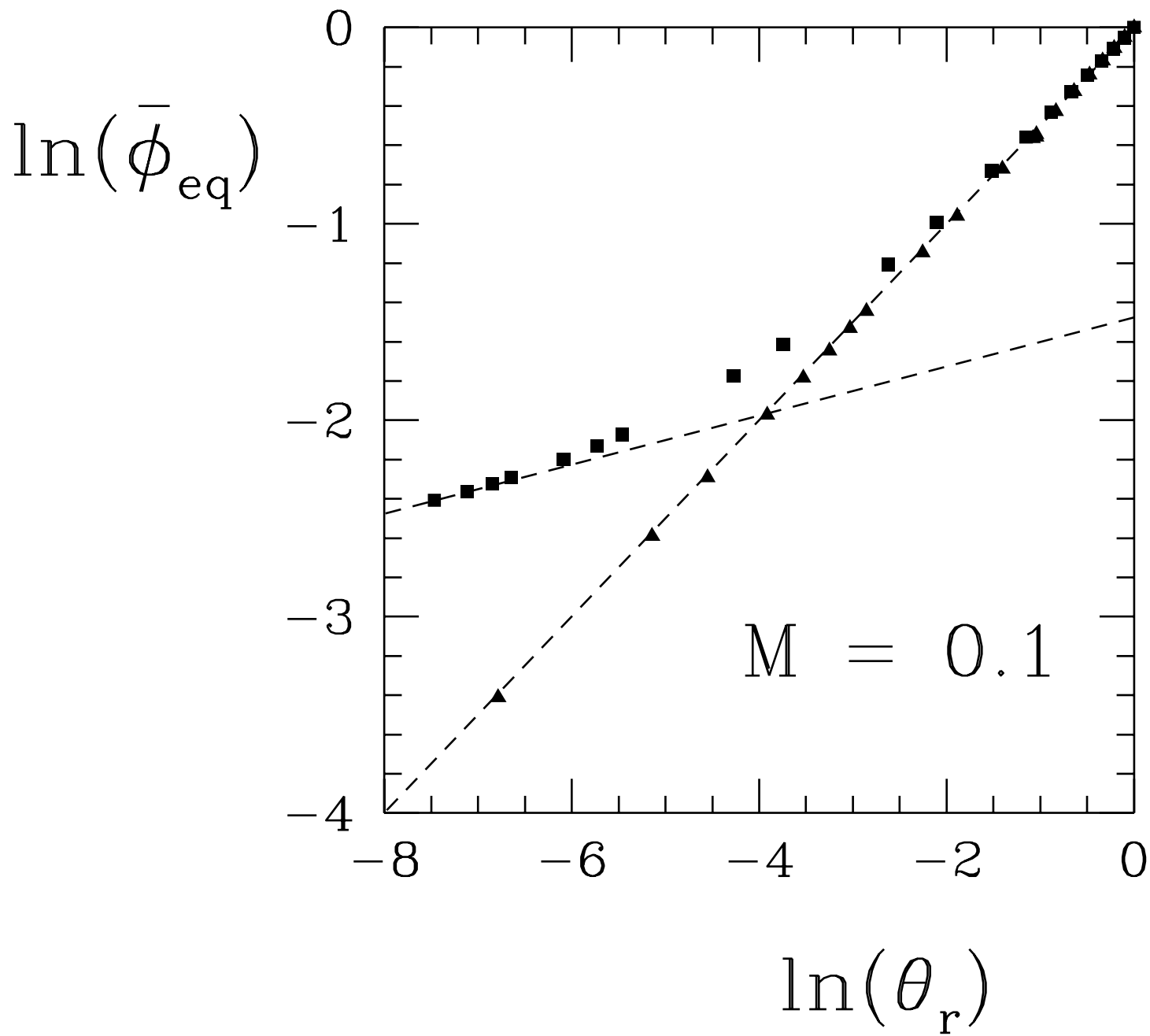
Figure 6. The variation in the equilibrium mean field
$\bar{\phi}_{\rm eq}$ with the reduced dimensionless temperature
$\theta_{r}$ from the lattice (squares) and the continuum
(triangles) for the temperature dependent potential. The dashed
lines have slopes of $1/8$ and $1/2$.\\

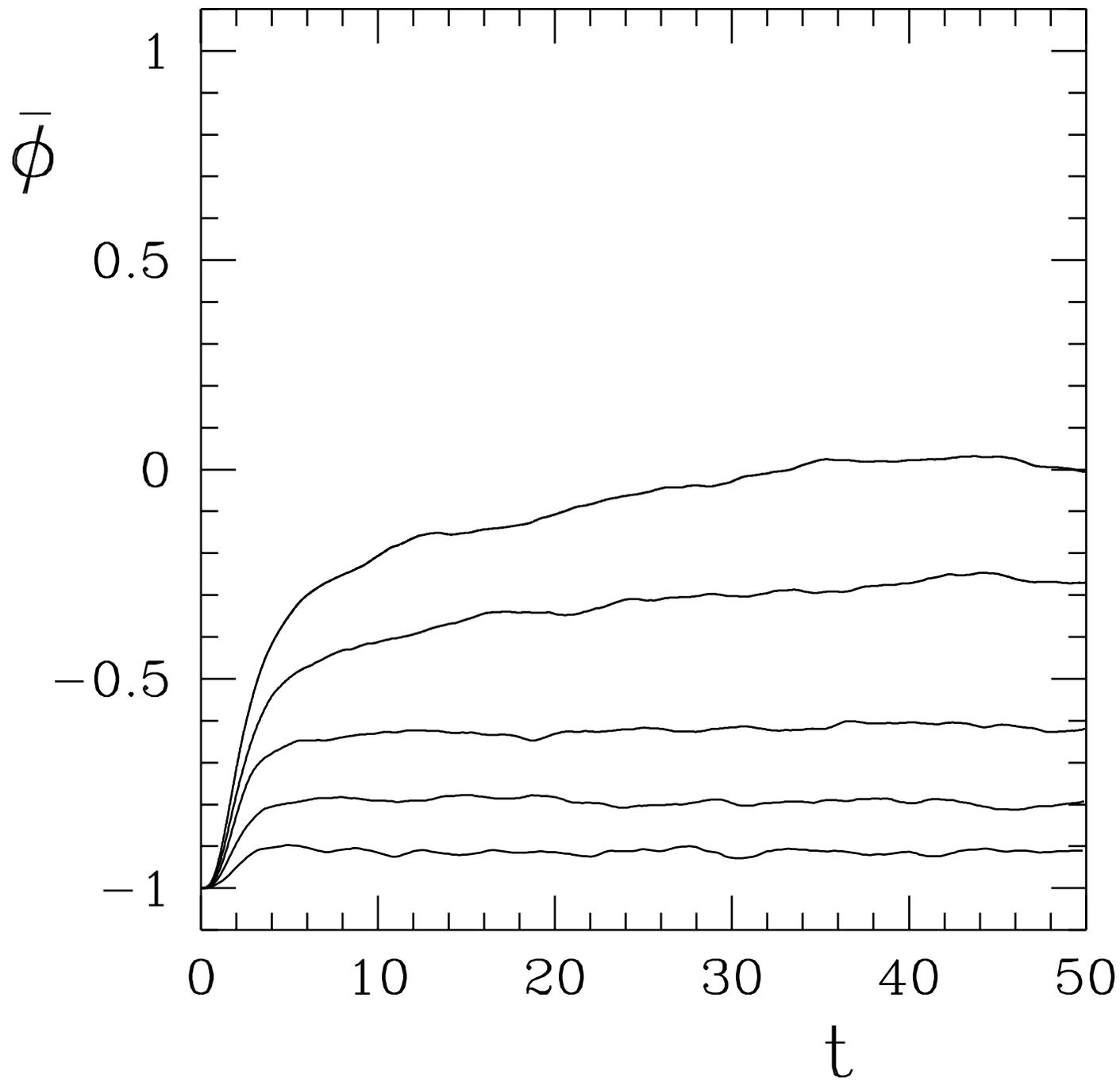
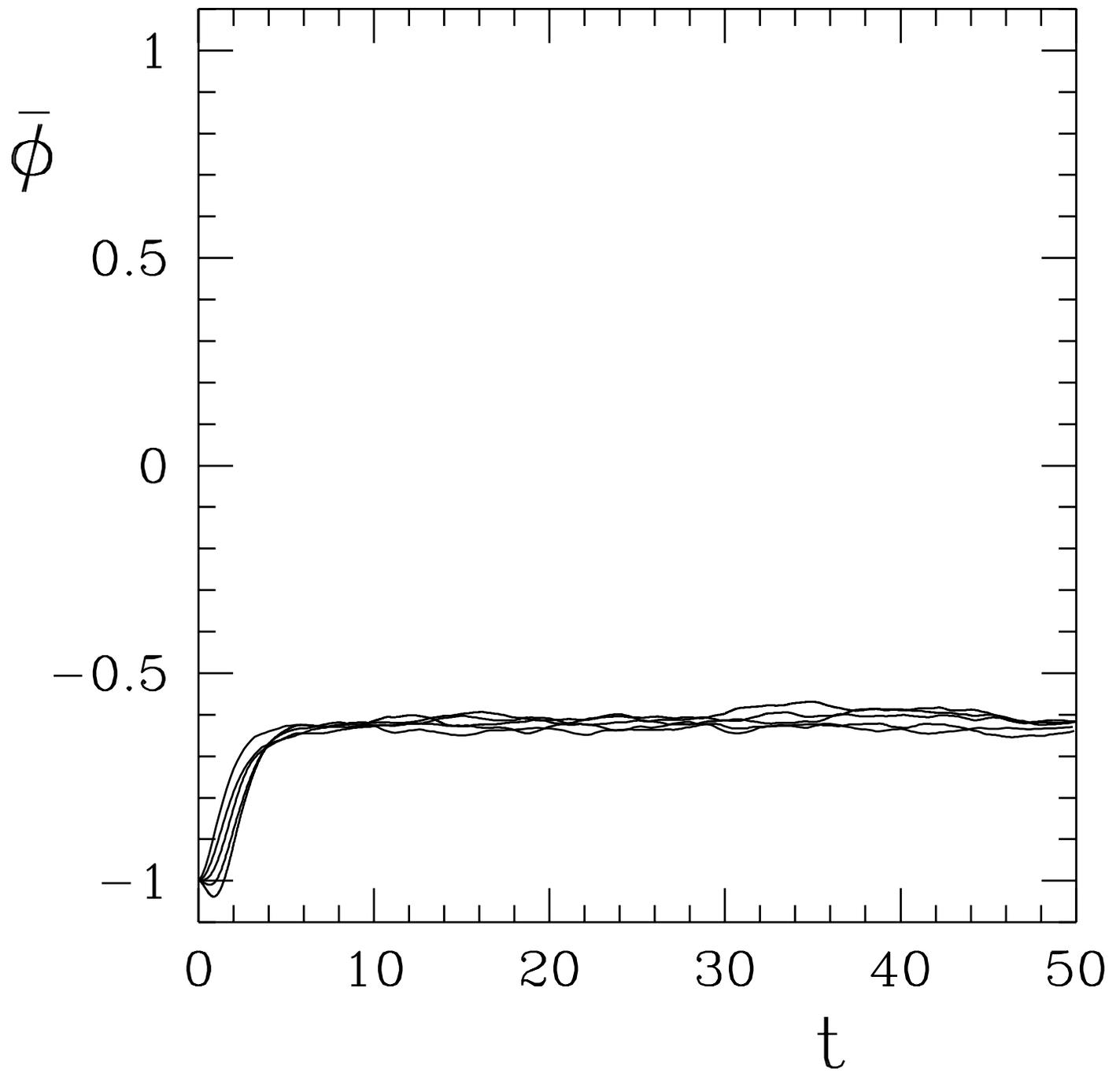
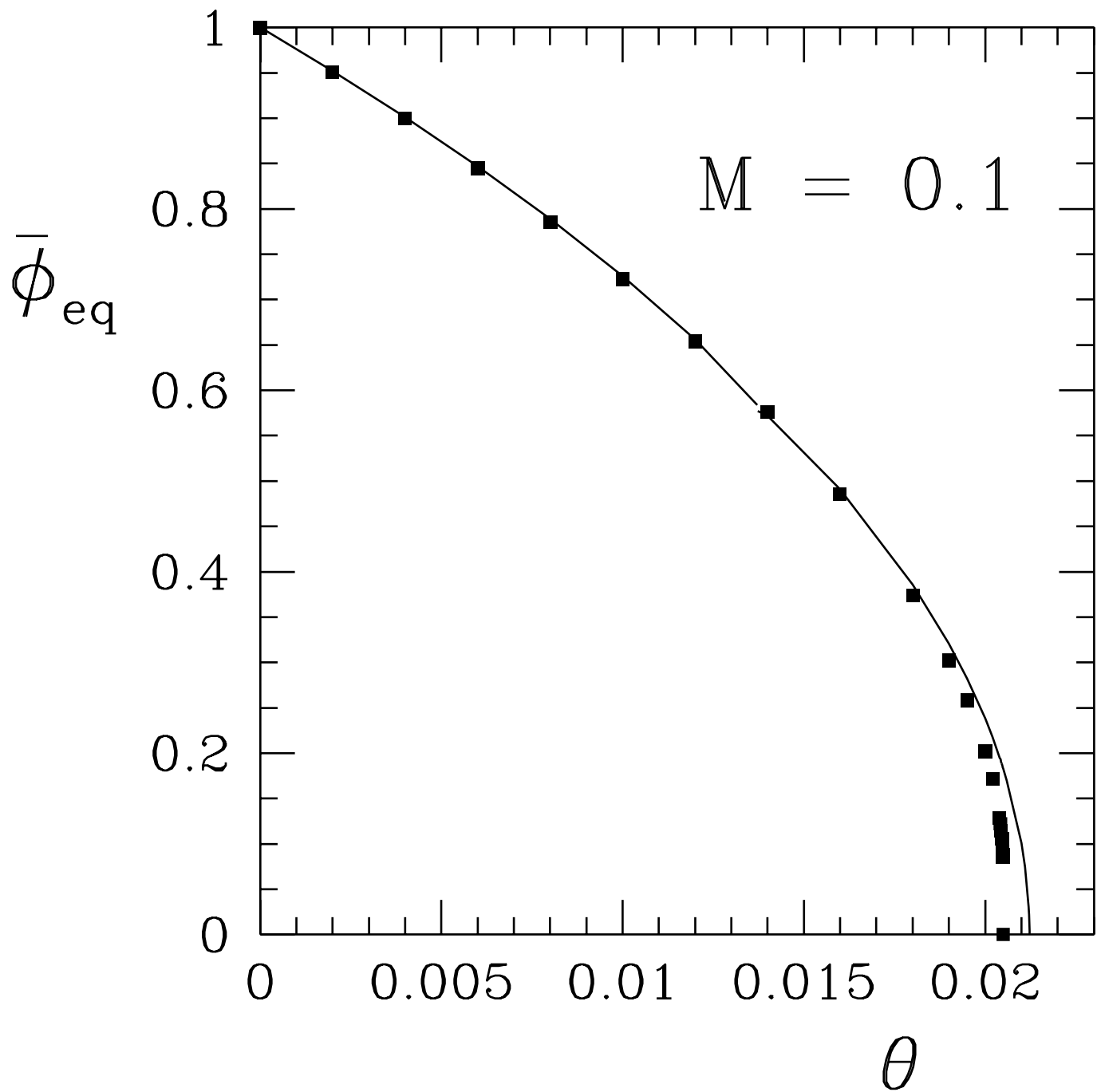
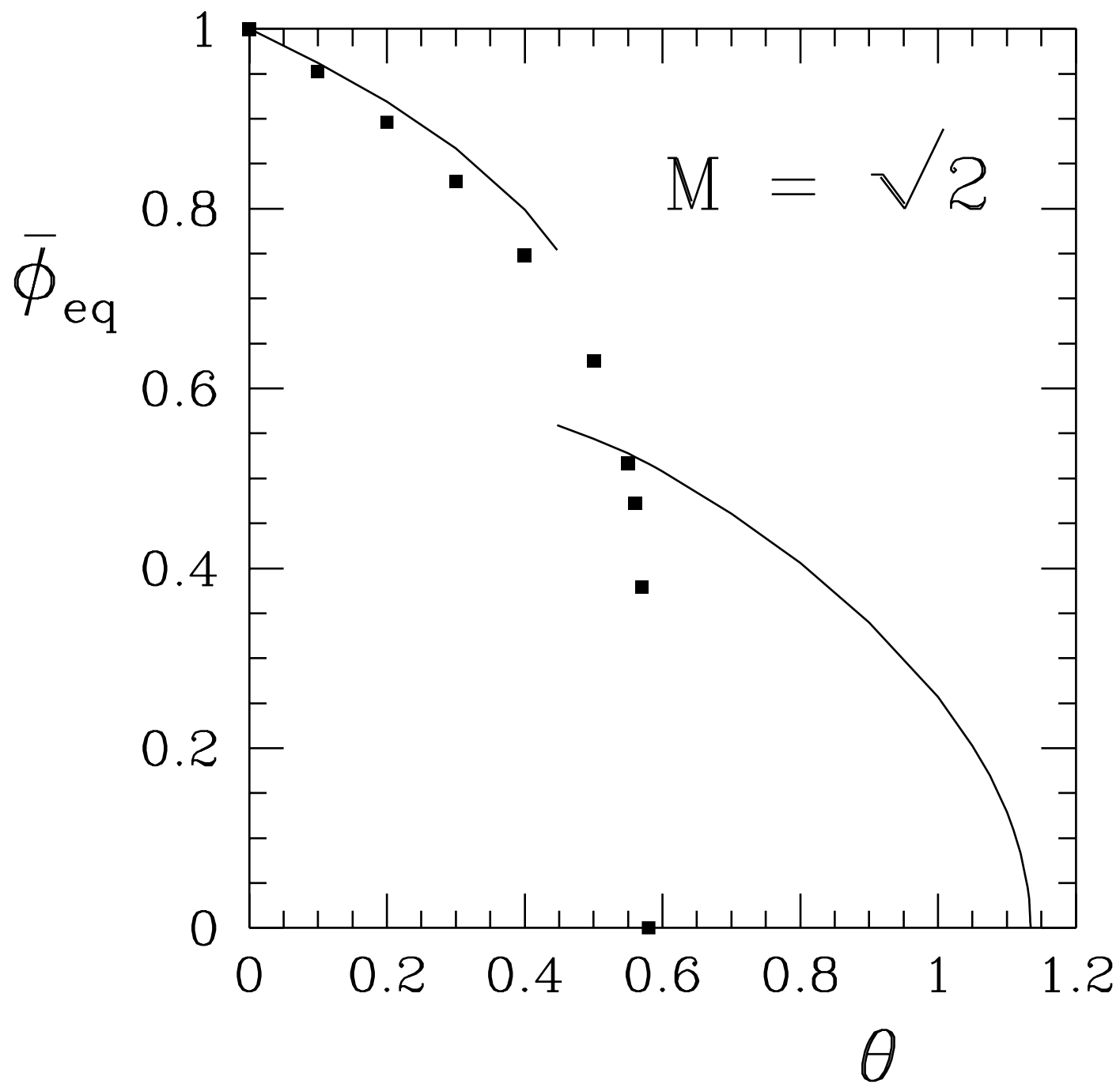
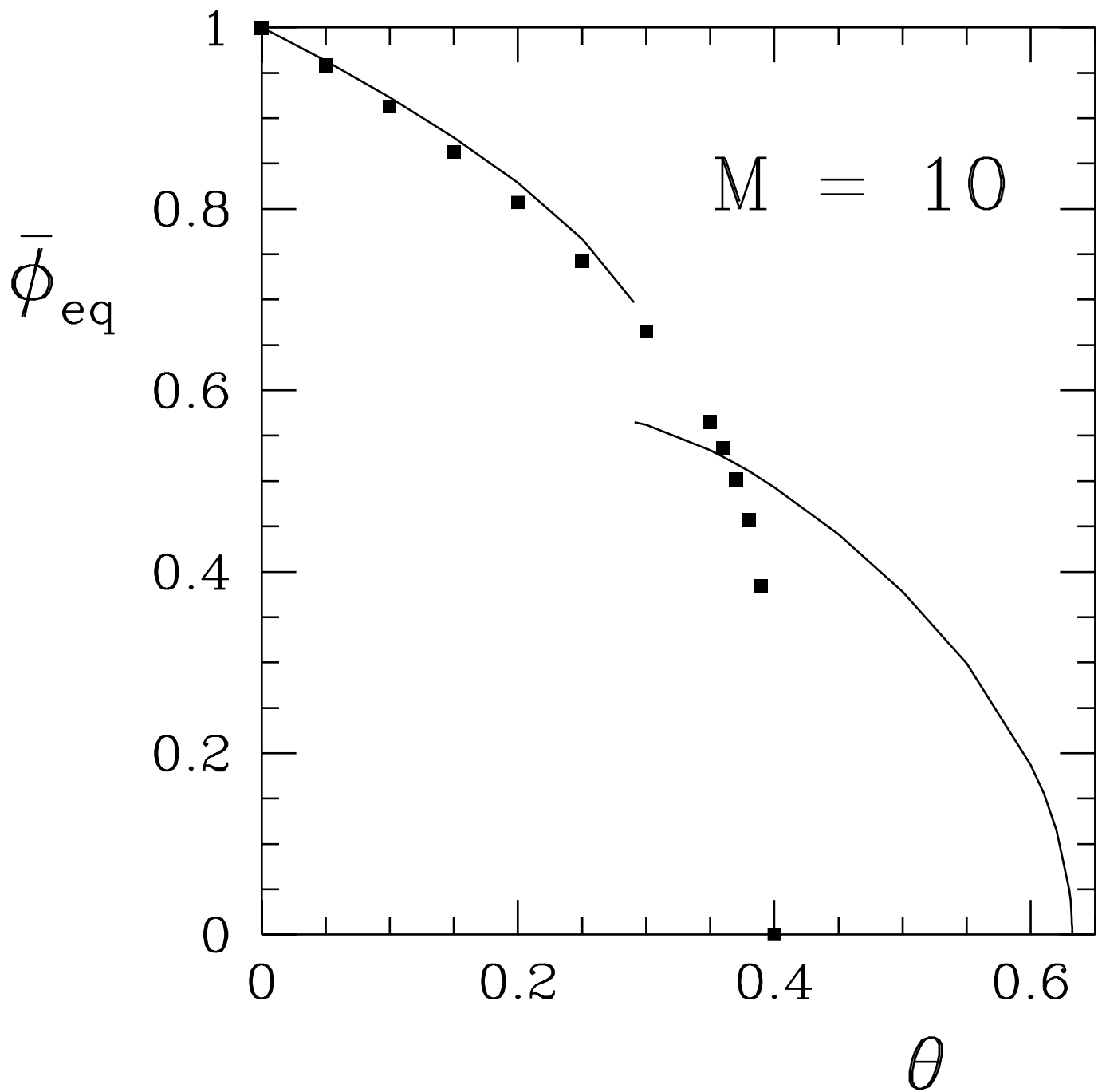
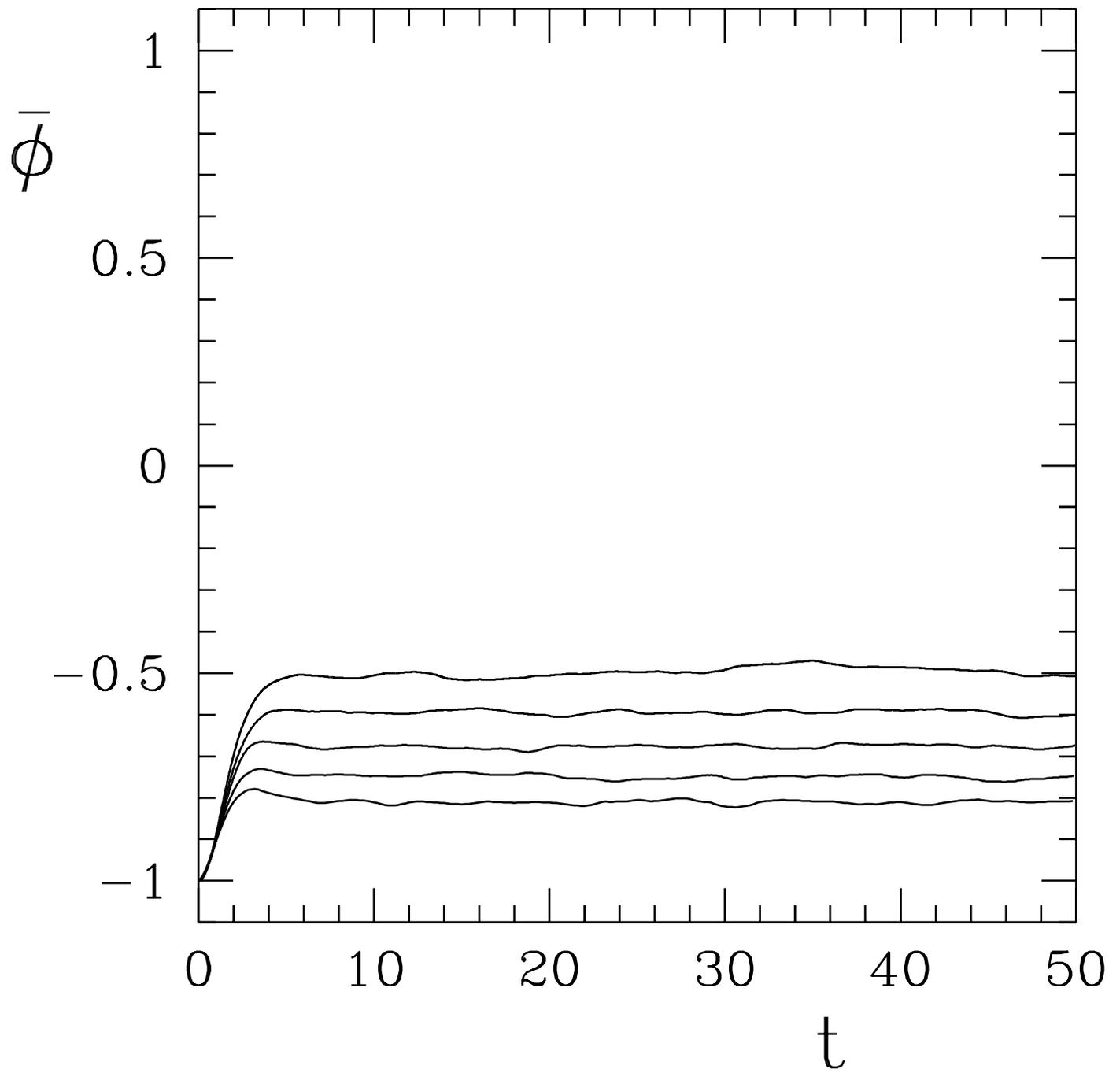
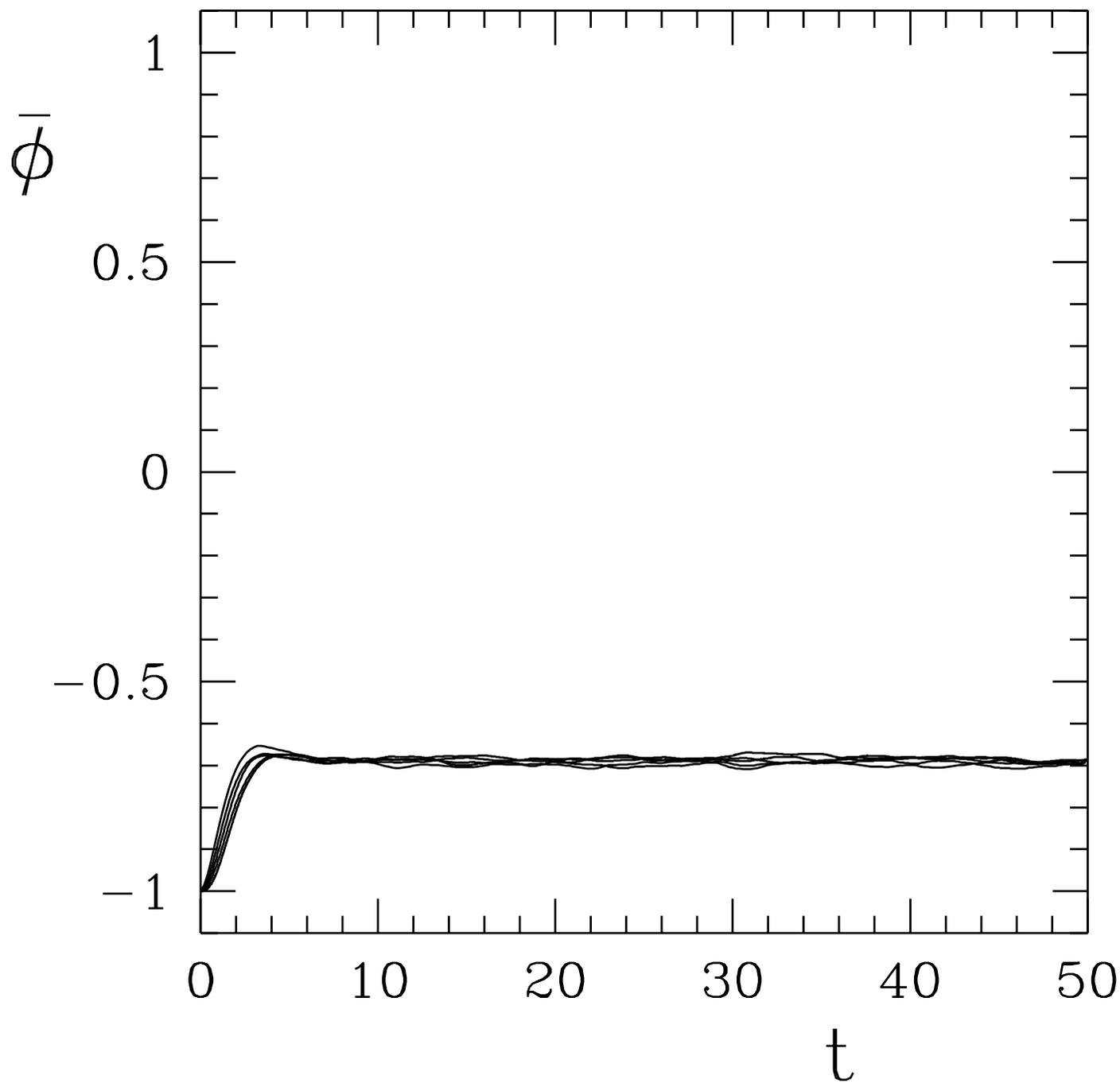
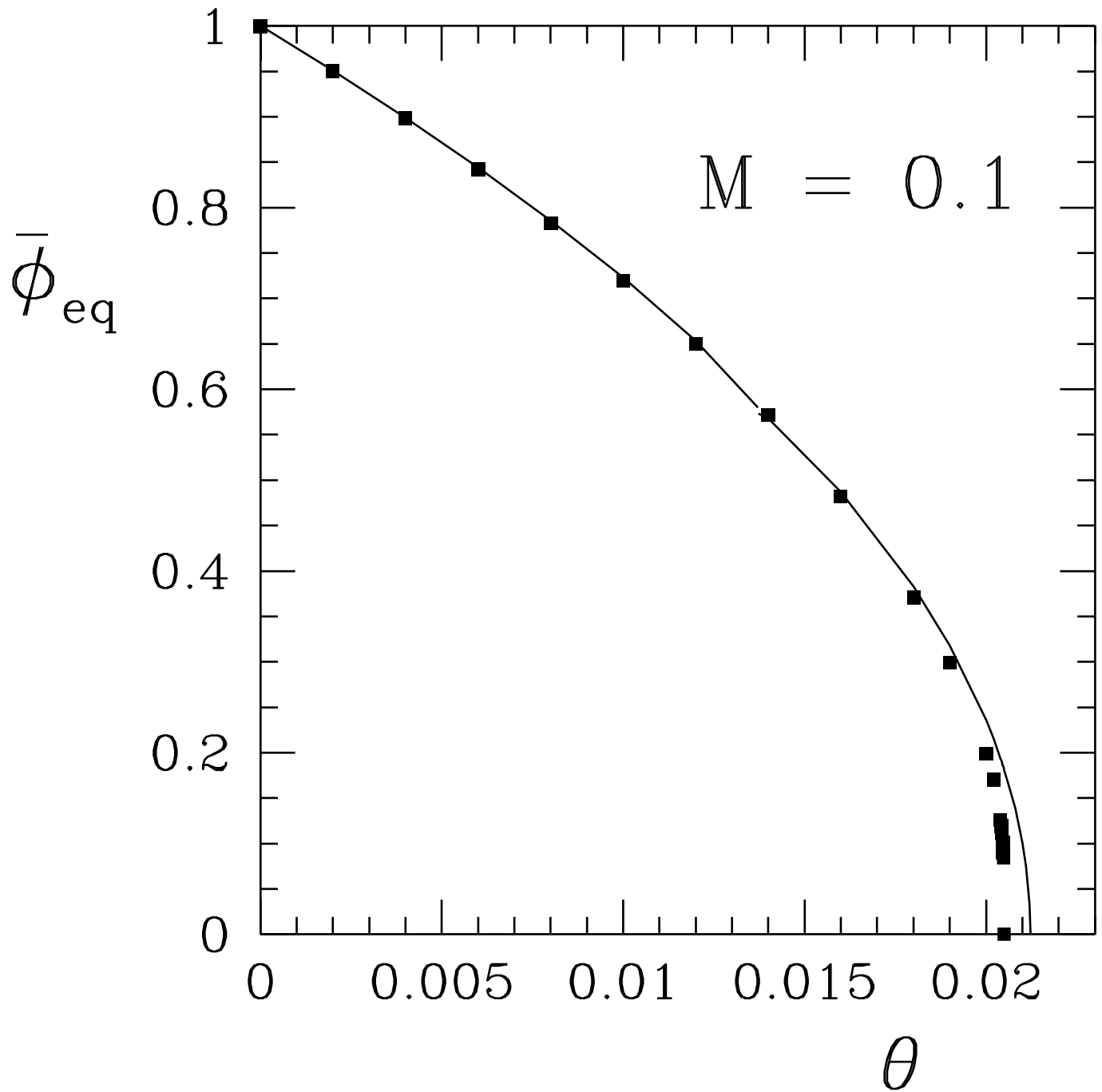
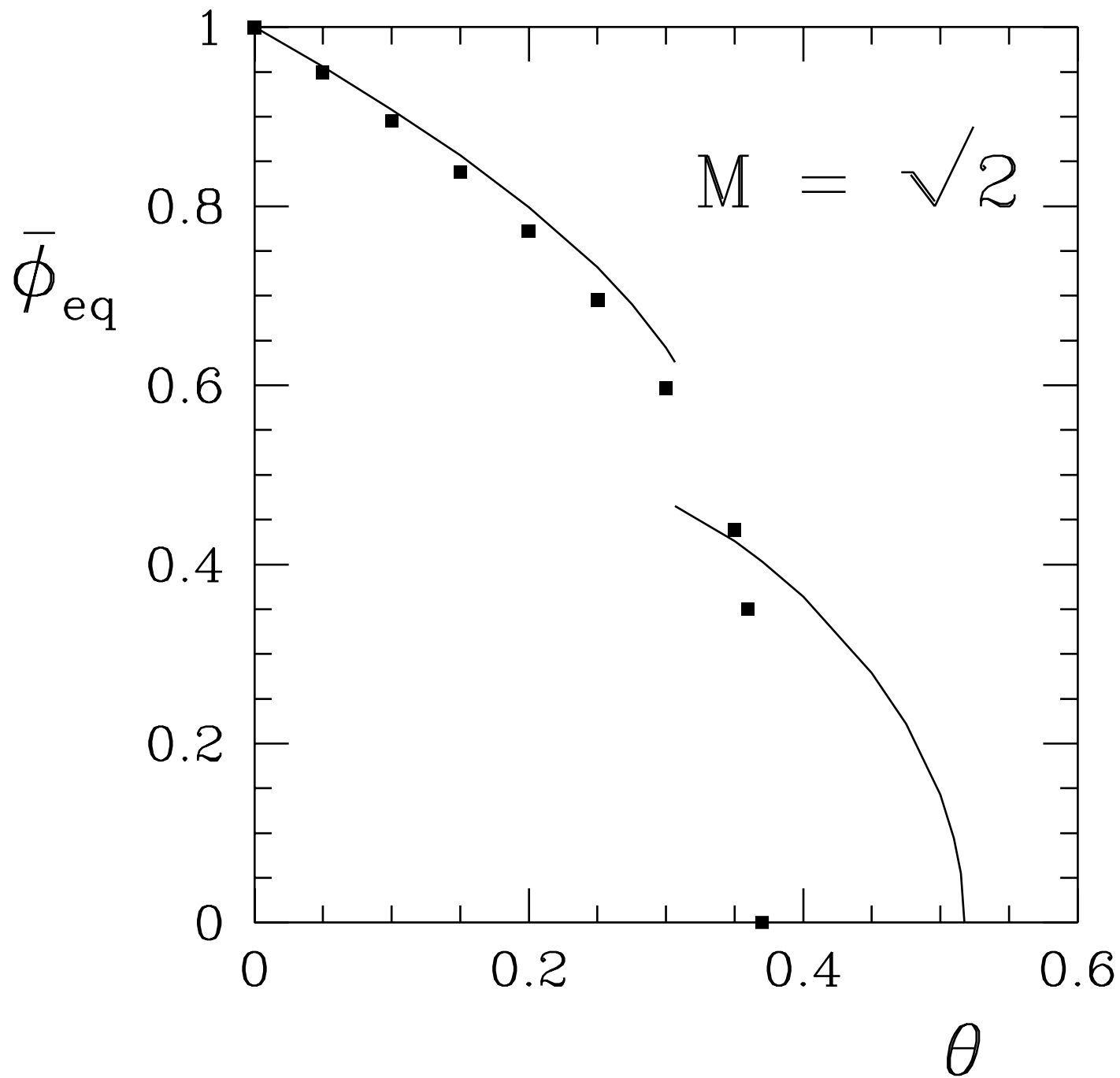
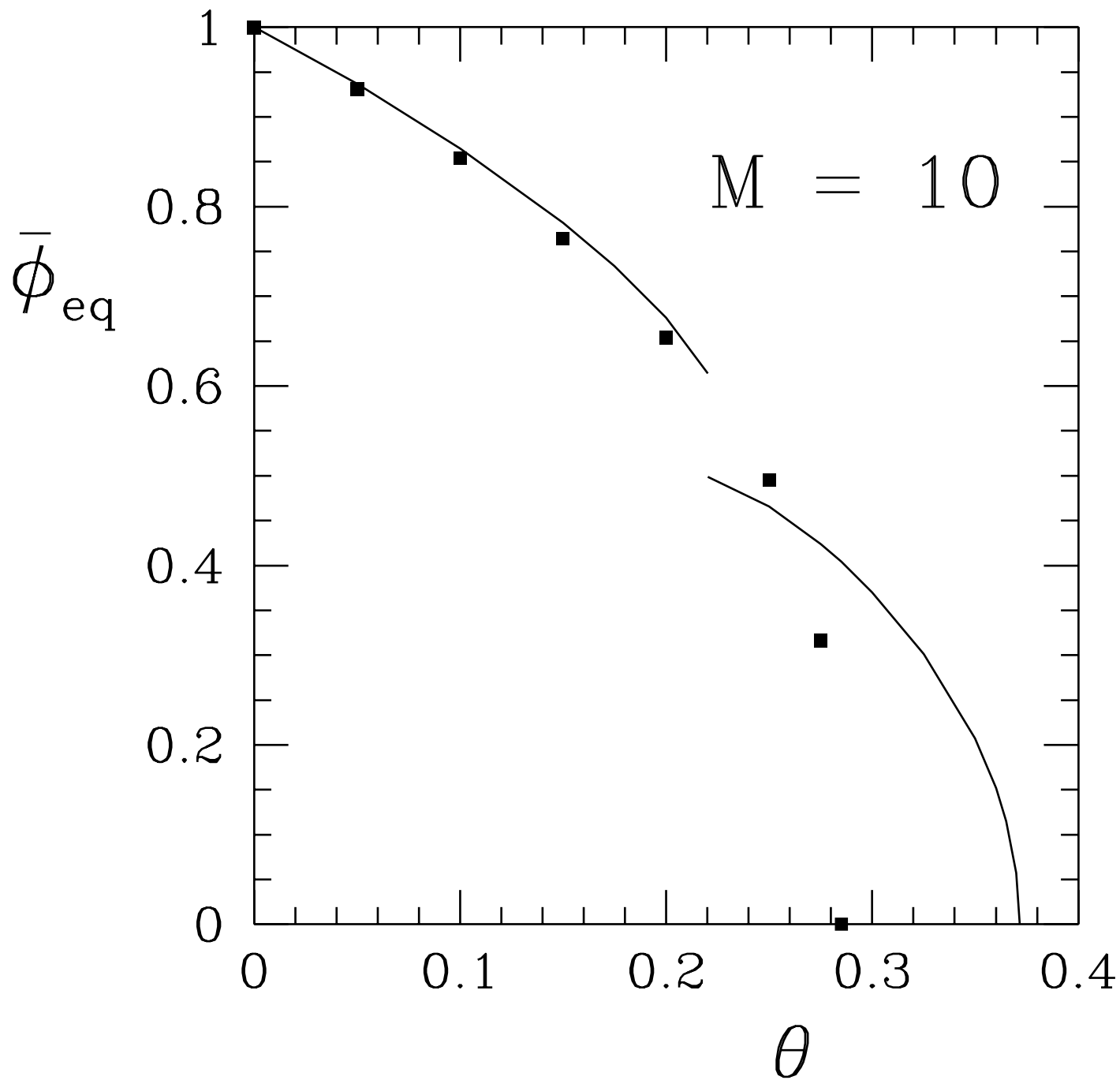
\end{document}